# Analysis, optimization, and characterization of magnetic photonic crystal structures and thin-film material layers

M. Vasiliev, K. Alameh, and M. Nur-E-Alam

Electron Science Research Institute, School of Science, Edith Cowan University, 270 Joondalup Dr, 6027, WA, Australia.

Corresponding author: Mikhail Vasiliev (m.vasiliev@ecu.edu.au)

**Abstract:**

MPC (Magneto-Photonic Crystal) Optimisation is a feature-rich Windows software application designed to enable researchers to analyze the optical and magneto-optical spectral properties of multilayers containing gyrotropic constituents. A set of computational algorithms aimed at enabling the design optimization and optical or magneto-optical (MO) spectral analysis of 1D magnetic photonic crystals (MPC) is reported, together with its Windows software implementation. Relevant material property datasets (e.g., the optical spectra of refractive index, absorption, and gyration) of several important optical and MO materials are included, enabling easy reproduction of the previously published results from the field of MPC-based Faraday rotator development, and an effective demonstration-quality introduction of future users to the multiple features of this package. We also report on the methods and algorithms used to obtain the absorption coefficient spectral dispersion datasets for new materials, for which the film thickness, transmission spectrum, and refractive index dispersion function are known.

**Keywords:** 1D magnetic photonic crystals; multilayer film modelling; modelling of Faraday rotation spectra; MPC optimization; exhaustive computation; materials characterization

**Program summary**

*Program Title:* Optimisation of 1D Magnetic Photonic Crystals (alternatively, MPC Optimisation)

*Program Installation Files:* available from
https://drive.google.com/open?id=1P3UgIu6nbfbmqXexrbppeiSIffplE-uv

*Licensing provisions:* Creative Commons Attribution-NonCommercial-3.0 Unported (CC BY-NC-3.0)

*Programming language*: Visual C#, compiled using Microsoft Visual Studio .NET 2003

*Nature of problem:* Calculation of the optical transmission, reflection, and Faraday rotation spectra in multilayer thin films containing gyrotropic constituents (magnetized material layers possessing magneto-optic properties); optimization of magnetic photonic crystal (MPC) designs aimed at achieving maximized transmission or reflection coincident with maximized Faraday rotation, according to sets of defined criteria; fitting of the experimentally measured transmission or reflection spectra to theory models; fitting of the absorption coefficient spectra of single-layer thin film materials using the data for optical transmission, film thickness, and refractive index spectra.

*Solution method:* The developed programme exhaustively calculates multiple possible multilayer structure designs, based on the design structure type(s) defined prior to running optimisation. Complex-valued 4×4 transfer matrix method (accounting for all dielectric tensor components, including the non-diagonal terms responsible for gyrotropic effects) is implemented to compute the complex field amplitudes and optical intensities in either the transmitted or reflected left-hand and right-hand circularly-polarized eigenwaves propagated through the thin-film-substrate structure.

*Restrictions:* The program is designed for use in conjunction with reliable optical constant datasets for up to 3 component dielectric materials, one of which can be modeled as magnetic dielectric possessing Faraday rotation; metallic layers are not implemented. Embedded active-X controls enabling graphical data output accept only up to 1000 data points per graphing control, whether plotting a single curve or several.

## 1. Introduction and background

In recent years, there has been some resurgence in the research interest dedicated to engineering and characterization of magneto-optic iron-garnet materials [1-4]. Thin-film magnetic garnets are semi-transparent magnetic dielectrics possessing record specific Faraday rotations of up to several thousand °/cm, in the near-infrared spectral range, if containing bismuth substitution [5]. Generically, the chemical composition of garnet materials of this type can be described by the formula $(Bi_xRe_{3-x})Fe_yM_{5-y}O_{12}$, where Re is rare-earth metal (e.g. Dy, Sm, Lu, Nd, or Ce), and M is transition metal such as Ga or Al [5]. The exploration of this important subclass of functional materials has decades-long history, starting from the days of bubble-domain magnetic memory development, and more recently, continued with renewed research activities, in application areas ranging from on-chip nonreciprocal components (waveguide isolators, [2, 6]), to magnetic recording [7].

Magnetic garnet materials synthesized by a range of thin-film deposition techniques and crystal-growth methods have also attracted a significant research and development momentum since 1990's, and throughout the last two decades, in areas ranging from photonic crystals to spintronics [8-14, 15]. Various approaches to the design and manufacture of magnetic photonic crystals (MPC) with tunable properties, and potentially suitable for the fabrication of non-reciprocal optical components have been reported [8-18]. Many research groups have focused on optimizing the thickness of one-dimensional (1D) MPC structures to simultaneously achieve 45° of Faraday rotation angle and maximum possible transmission at optical telecommunication wavelengths. MPCs based on quarter-wavelength thin-film stacks, which are sequences of magnetic and nonmagnetic layers with multiple embedded phase shifts (termed defects, or missing layers), have been shown to possess a significant potential for practical implementation of integrated optical isolators. This is due to the necessity of approaching Faraday rotations as large as 45°, which has been shown to be attainable, due to the resonant enhancement of Faraday rotation observed in such structures. Complex 1D MPC designs featuring "flat-top" spectral response, with almost 100% transmission within a large bandwidth (several nanometers), and close to 45° of Faraday rotation at 1550 nm can contain in excess of 200 layers [10,11], limiting their applications to the near-infrared range, where the magnetic garnets possess very low (negligible) optical absorption. Since the original development of this present MPC Optimisation program in 2005 [18], multiple garnet materials development efforts have been undertaken within our group [19-21], all of which have relied substantially on material characterisation options featured within the same software package. In particular, the option of deriving the data for the spectral dispersion of the absorption coefficient using the measured transmission spectrum and refractive index dispersion data has been very useful in characterisation of new nanocomposite-type garnet materials synthesized by co-sputtering using an extra oxide material source [19]. The present-day performance limits of 1D MPC in the visible spectral range have been evaluated using the same software [22] and using the available spectral data on the optical properties of the best-performing magnetic garnet compositions. Optical constants data of multiple recently-synthesized magnetic garnet compositions have been evaluated using the measured transmission spectra in conjunction with MPC Optimisation software and Swanepoel envelope method [23]. Our group's preferred method for calibrating the quartz microbalance sensor's tooling factors of various deposition sources also involves fitting the actual film layer thickness using MPC Optimisation software in conjunction with measured transmission spectra.

A graphical snapshot of the controls and features implemented within MPC Optimisation software is shown in Figure 1, which also presents a sample optimisation result. The computation time to obtain this result is typically around 1 minute (after analyzing almost 5000 MPC designs out of possible 10000 defined by the pre-set film structure features; some of calculated designs exceed the maximum limits set for thickness or layer number, and thus are not fully evaluated). The default spectral range of calculations is extremely broad, slowing down the optimisation algorithm significantly, since the default wavelength settings and resolution have been entered to assist in fitting film layer thicknesses conveniently, which is one of the most common everyday applications of this software. Running actual MPC optimisation algorithm is best performed by zooming onto the spectral range of interest, which is usually represented by a wavelength region surrounding a narrow peak of transmission or reflection, where it is possible to engineer the enhancement of Faraday rotation.

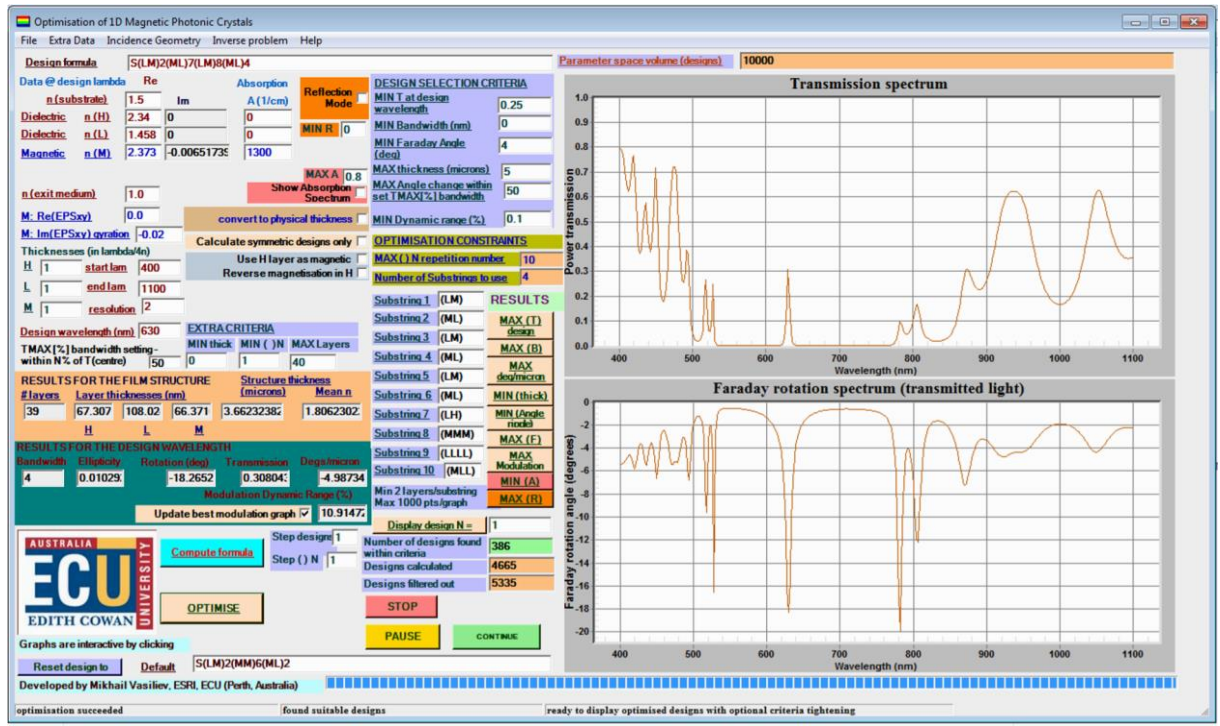

**Figure 1.** Front-panel controls of MPC Optimisation software and a sample optimised MPC design obtained by clicking the "Optimise" button without changing default-entered data. The result is MPC structure designed to operate at 630 nm, reliant on a magnetic garnet material of specific Faraday rotation near 2°/μm at that wavelength, however, the structure enhances the Faraday rotation to 4.98°/μm within the spectral transmission peak.

The default-entered materials-related data used to optimise MPC designs (within the parameter space and constraints also entered as default values not relevant to any particular application) relate to a MPC design based on magnetic garnet composition $Bi_2Dy_1Fe_4Ga_1O_{12}$ and $SiO_2$ L-type layers, deposited onto a glass substrate. The optical constants and gyration data for wavelengths near 630 nm have been obtained from thick garnet layers of this composition, and the corresponding index dispersion dataset is also pre-loaded into the menu item "Extra Data | Account for refractive index dispersion".

The contents of the compiled html (.chm) Help file accessible from Help menu are sufficient to enable most beginner-level MPC (or thin-film) designers to quickly learn the main features of program and its data representation formats. The following sections of present article describe these main features in detail, with examples provided to enable the productive and convenient use of this feature-rich software package.

## 2. Overview of package operation and key examples

Since the structure of 1D MPCs is essentially represented by a sequence of magnetic and non-magnetic material layers on a substrate, most of the core terminology, design structure description conventions, and analysis techniques are derived from the field of multilayer thin-film design. The input datasets necessary to define the layer sequence and the optical properties of each material type and individual layer are entered into the relevant textboxes, starting from the top-left corner of the Windows Form. The **Design formula** field defines the layer sequence, starting necessarily from the capital letter S defining the substrate. The layer sequence must be entered in an alphanumeric string format containing special symbols such as round, square, or curled brackets, corresponding to one of the three main design-string representation types. These types are: (i) quarter 4 wave-stack based notation, e.g. SLH(ML)2 which is perhaps the most common in thin-film design; (ii) physical thickness-based notation, e.g. SM[1000]L[50], and (iii) advanced designs notation, e.g. S{1.0}(L/2HL/2)1{1.06}(L/2HL/2)3 further described within help documentation. The optical properties of the substrate material (which is presumed to be dielectric, semi-infinite, and non-absorbing) are defined by only the real part of refractive index. Up to three different optical materials are allowed, denoted by letters H, L, and M, however, H-type layers are not restricted to mean "high refractive index"; the layer naming conventions are derived from thin-film terminology for convenience only. M layers can optionally be modelled as magnetic dielectrics, in which case gyration value at the design wavelength (the imaginary part of the non-diagonal dielectric tensor component, or Im($\varepsilon_{xy}$)) needs to be entered into its relevant textbox. This imaginary component of the off-diagonal dielectric tensor element describes magnetic circular birefringence, manifesting as Faraday rotation of the polarisation plane in the transmitted or reflected light waves. The real part of this tensor component can be treated as zero, in most applications, unless the experimentally-measured value is known, which is related to characterising magnetic circular dichroism and polarisation state ellipticity effects. For applications requiring good numerical accuracy over broad spectral ranges, the spectral dispersion of both the gyration and the refractive indices of all relevant materials need to be loaded from .txt formatted data files, using the options within Extra Data menu. Optionally, H-type layers can be selected to also represent a magnetic dielectric, with the same off-diagonal dielectric tensor components as in M-type material (but optionally with different refractive index and absorption), in order to model a special physical situation in which an MPC has layer-specific magnetization reversal possibilities in some individual magnetic layers within structure. If M layers are modeled as non-magnetic (e.g. just implying medium-index dielectric layers), then both the real and imaginary parts of $\varepsilon_{xy}$ can be entered as zero values; however if a small gyration is still entered by error, it will not measurably distort the transmission or reflection spectral calculations; the results regarding the Faraday rotation spectra should in this case be ignored.

The details of the physical situation being modeled, in terms of incidence geometry, and the ways in which the transmitted (or reflected) light intensity is normalized with respect to the incident wave intensity, are defined using sets of checkboxes within a Menu entry "Incidence Geometry", where the relevant descriptions are given. The incidence geometry settings defined as default are the ones used most often and generic in general, since these enable the convenient and accurate fitting of lab-measured transmission spectra in deposited film-substrate systems, to the corresponding theoretically-modeled spectra, where the effects of each interface (including the back surface of substrate) are correctly accounted for in the model. The only parameter not accounted for is the physical thickness of substrate, which is modeled as non-absorbing. Alternative settings for the incidence geometry are useful for considering more theoretical situations, such as calculations of the optical intensity transmitted into a semi-infinite substrate medium, or in reflection-mode calculations, where it is often necessary to compare reflected-wave intensities, which vary with the direction of incidence.

One of the most important material parameters in all layer types is the optical absorption coefficient at the design wavelength (entered in cm$^{-1}$ into relevant textboxes; the corresponding extinction coefficients will then be displayed after the film design is characterised by pressing the Compute formula button), especially for materials possessing significant spectral dependency in their optical absorption. For accurate characterisation of thin films or MPCs, ideally, every material should have its optical constants dataset available for loading from the Extra Data menu option "Account for refractive index dispersion" and loaded into the specialised Form (shown in Fig. 2) prior to calculations. Material-specific optical constants data files can be prepared using zeros entered in place of an unknown absorption coefficient, for the purposes of physical layer thickness fitting, based
on the accurately measured transmission spectrum data (as discussed more in detail in the subsequent sections). The end-of-file (EOF) marker in these .txt material data files prepared using editor applications such as Notepad must be placed immediately after the last figure in the last column, by way of making sure to delete any possible symbols or empty rows space below.

**Figure 2.** Form dedicated to loading the optical constants data from text files prepared as shown, using the software-specific 27-row wavelength grid and containing the columns data for the refractive index and optical absorption coefficient (in cm$^{-1}$) at each wavelength point within the spectral grid.

Once the refractive index dispersion information is loaded from data file(s), the data in textboxes corresponding to the index and absorption at the design wavelength are no longer used in main spectral calculations, but used only for calculating the quarter-wave stack physical thickness in nm. For all wavelength points in-between the 27-point data grid, the values of refractive index and absorption are linearly interpolated from the nearest grid-located points. This can erroneously cause small spectral shifts appearing in the spectral locations of the transmission, reflection, or Faraday rotation peaks, seen away from the precise design wavelength, if the interpolated refractive index at that wavelength does not coincide with the value entered into front-panel textbox. This is not a significant issue for the experienced designers of MPC, once the origin of these possible small data errors is known or eliminated by entering precise (same as interpolated from the index dataset for the design wavelength) index data into front-panel textboxes. It is known apriori that the actual spectral response peak locations in quarter-wave stack-type designs will appear precisely at the design wavelengths, due to the nature of optical interference-related phenomena. In situations when the refractive index dispersion data are only available within a limited spectral interval of interest, rather than for all wavelengths in the 350–1600 nm range, it is recommended to enter the available refractive index data into the newly-generated data file. The refractive index and absorption coefficient values at other wavelengths still need to be entered. The recommended practice is as follows: if, for example, the available data starts from 500 nm, enter the same values as are known at 500 nm into the wavelength grid positions for all shorter wavelengths; alternatively use any available models to predict the unknown values (e.g. Cauchy dispersion model). If the refractive indices or absorption are only available up to 800 nm, it is best to enter (for all larger wavelengths), the same (n, A) data as at the last data point (800 nm).

The refractive index of the exit medium surrounding the substrate-film system is defined only by its real part (typical value is 1.0 for air), and the same exit medium is presumed to precede the (thickness-undefined) substrate, and to exist beyond the last deposited film layer, regardless of the direction of light incidence. The checkbox "Reflection Mode" sets up the calculations of the reflection spectra, and also the Faraday rotation spectrum for the reflected light, if checked. Optionally, the Show Absorption Spectrum checkbox can be checked, after which the Compute formula button will initiate new calculations, resulting in the display of the calculated absorption spectrum.

*2.1. Multi-defect multilayer MPC characterisation and optimisation examples suitable for validating calculations*

To illustrate the suitability of software to correctly calculate the spectral responses of complex, multi-defect MPC designs, it is easiest to use the design or optimisation examples described in the published literature, e.g. [11] and [22]. One of the common goals of optimising the MPC structures has been to achieve strong spectral peaks in either the transmission or reflection (ideally approaching 100%), coincident spectrally with peaks of enhanced Faraday rotation, in either the transmission or reflection-mode operation, and ideally approaching 45° - if efficient modulation of light intensity is required. It is important to note that Faraday rotation in the reflected wave is different in its physical nature from Kerr rotation [5] (even though there may exist some terminological misinterpretations, even in the published literature); this program calculates only the angles of polarisation-plane rotation due to Faraday effect, in either geometry, and does not account for Kerr effect. Figure 3 presents a graphical summary of the input parameters needed to be entered into relevant textboxes to evaluate one remarkable MPC design from published literature [11], as well as the results of calculations. The practical implementation of this MPC design can be expected to be difficult, due to factors such as extremely high number of layers, large total thickness, the expected non-zero (but possibly well below about 0.1 cm$^{-1}$) optical absorption coefficient in garnets at 1550 nm, as well as scattering effects expected to occur at multiple layer boundaries. From the theoretical insight perspective, this high-performance MPC design is still an outstanding example of MPC application potential, especially in systems using optimised surrounding media, index-matched to the mean refractive index of structure. A .mpc file (MPC Design from JLT Vol. 19 No. 12 p. 1964.mpc) is included in the subfolder "Optimisation results files" of the program installation directory, and can be loaded from the File menu, enabling easy re-calculation of the contents of Fig. 3, using pre-loaded design data. The way this MPC has been modelled also involves the customised Incidence Geometry settings, where film-side incidence is modeled, without normalising the transmitted intensity after the substrate. Rather, transmission into the substrate material is modelled, and this is why the modeled transmission within 1550 nm peak closely approaches 100%. Running MPC Optimisation algorithm, on the other hand, requires using substrate-side incidence geometry settings only.

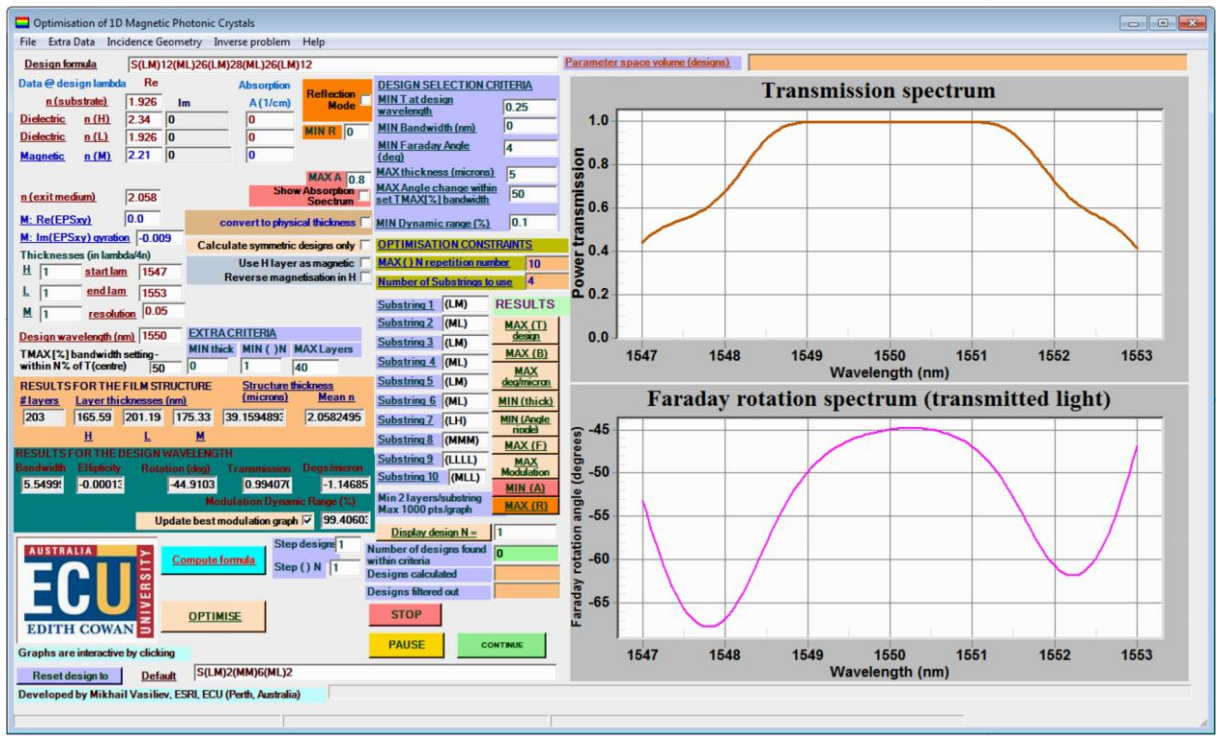

**Fig. 3.** MPC optimisation software used to reproduce the flat-top MPC transmission and Faraday rotation spectral properties for a 4defect design reported in [11]. The calculated graphs (obtained after entering the design data and pressing the Compute formula button) reproduce the data originally presented graphically in Fig. 3(b) of Ref. 11.

All materials-related data values were used as per description in Ref. 11, and the data in the calculated graphs reproduce the results presented in Fig. 3(b) in Ref. 11 with precision.

Due to materials-related constraints, such as the spectral dependencies of the absorption coefficient and gyration, achieving strong enhancements in Faraday rotation simultaneously with low optical losses becomes progressively more difficult with the reducing design wavelength. Across the visible spectral range, the optical absorption in bismuth-substituted iron garnet materials becomes the limiting factor, placing stringent limits on the achievable MPC performance characteristics, regardless of their intended application area or the design type. Therefore, the ability to generate and compare multiple and differently-structured optimised MPC designs is crucial for achieving the best possible performance characteristics, limited only by the fundamental, materials-related constraints.

In order to directly reproduce the optimisation result reported in [11], by way of running constrained optimisation algorithm using the MPC Optimisation software, the materials-related datasets and a set of optimisation constraints as shown in Figure 4 must be used prior to clicking "Optimise" button. It is also necessary to set the substrate-side incidence geometry, and uncheck the second checkbox related to the way the transmitted intensity is normalised. The "flat-top" optimised MPC design will be retrieved from the database of calculated MPC designs within the set of defined criteria (60 different designs will be found to fit these overall criteria and constraints, as set per data of Fig. 4), by selecting the design with maximised spectral response bandwidth. This is done by pressing MAX (B) button. Importantly, all design substrings must be entered as per Fig. 4 to reproduce this optimisation result, with precisely 5 substrings and maximum 30 entered for the substring-repetition index. Another constraint which needs to be entered relates to the maximum allowed layer number being 203; checkbox "Calculate symmetric designs only" needs to be checked, and it is best to set the spectral resolution to 1 nm during optimisation, followed by changing it to 0.1 nm for calculating the spectral properties more accurately. The optimised design equivalent to the design of Fig. 3 (and Ref. 11) is retrieved from 60 possible designs found within criteria, by pressing the MAX (B) button. This retrieves the design with maximised full width at half maximum (FWHM) bandwidth, according to the data entered into the TMAX(%) textbox.

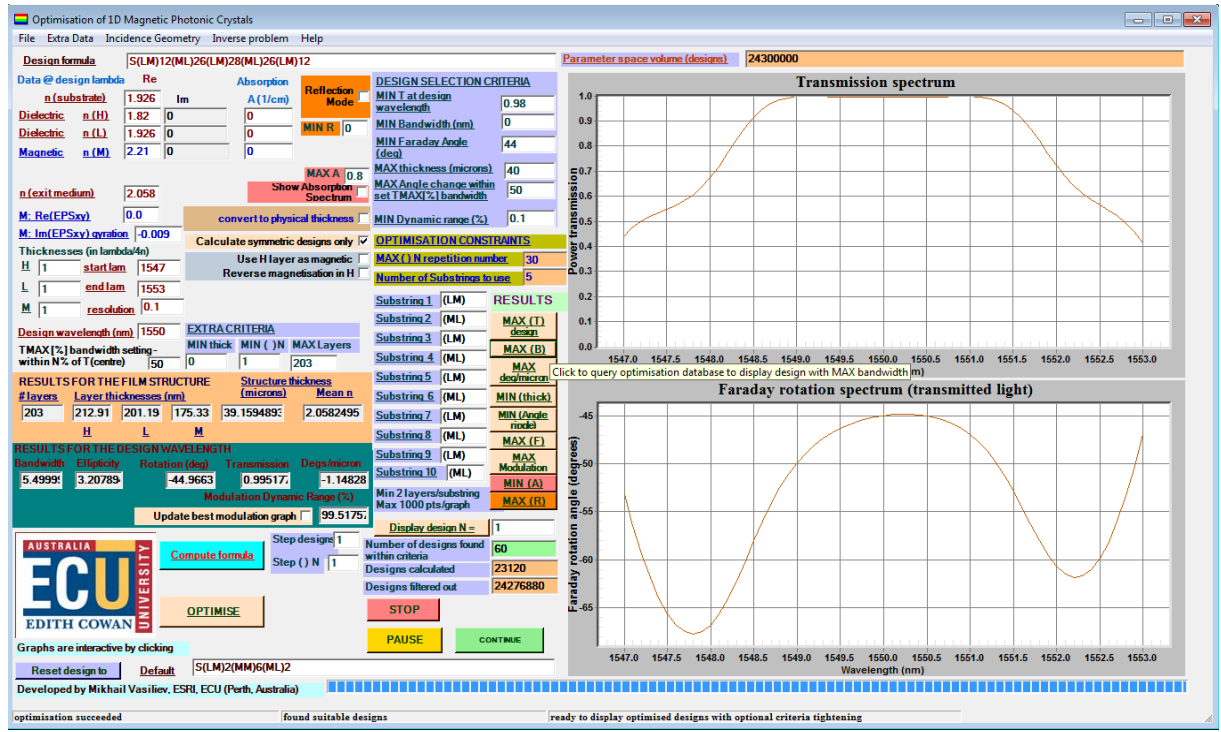

**Fig. 4.** Optimised "Flat-top" high-performance MPC design reproduced by running MPC Optimisation algorithm, using checkbox "Calculate symmetric designs only" for speeding up calculations. The exact 4-defect MPC design reported in [11] (Fig. 3(b) of Ref. 11) is shown after selecting the design with maximum spectral response bandwidth from the 60 possible MPC designs found to be within optimisation criteria and constrains, as shown.

The settings applied to the normalisation of the transmitted intensity within Incidence Geometry menu corresponded to the physical situation equivalent to that applied during the calculation of MPC properties in Ref. 11, stipulating the transmitted intensity normalisation procedure involving transmission "from within" the substrate material, into the index-matched exit medium.

Figure 5 shows a graphic summary of MPC optimisation results first reported in [22], which illustrate the current performance limits of MPC designs aimed at developing transmission-mode magneto-optic light intensity modulators working at near 650 nm. The optical and MO material parameters relevant to $Bi_2Dy_1Fe_4Ga_1O_{12}$ and similar highly bismuth-substituted nanocrystalline garnet materials (synthesized by techniques such as RF sputtering) were used in the calculations. The graphical information is reproduced from Ref. 22.

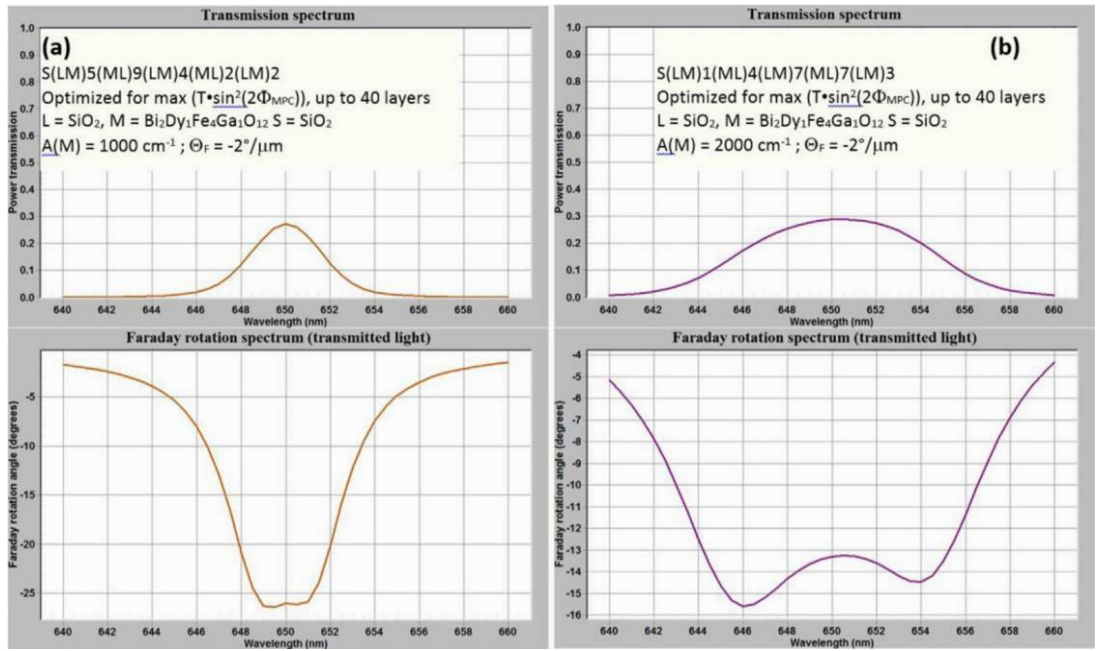

**Fig. 5.** Calculated spectral performance parameters for multi-defect (4-defect structures, having up to 40 total layers) MPCs optimized by exhaustive computation to achieve a maximum light intensity modulation capability (maximized value of parameter (T· $sin^2(2\Phi_{MPC})$)), for n(M) = 2.376, n(L) = 1.458, and using two different optical absorption coefficients for magnetic material at 650 nm (a) $\alpha$ = 1000 cm$^{-1}$; (b) $\alpha$ = 2000 cm$^{-1}$. These refractive index and absorption coefficients were considered constant within the design-specific wavelength region, as well as gyration (-0.02), corresponding to approximately 2 °/µm near 650 nm.

The data of Fig. 5 can be reproduced by running optimisation of 4-defect structures with up to 40 layers and thickness up to 5 µm, composed of five sequenced (LM) and (ML) substrings, as shown, using default entered n(L) value and n(M) = 2.376. The gyration value corresponding to 2 °/µm needs to be entered as -0.02 for wavelengths near 650 nm, accounting for the composition-specific sign of Faraday rotation, which is defined by convention reported in [5] and other sources. Maximum repetition index value can be set to either 10 or e.g. 12, prior to running the optimisation with either absorption coefficient.

The results shown in Fig. 5 illustrate clearly that the optical absorption is the limiting factor in visible-range MPC design, even at long-wavelength red wavelengths, where the absorption is still moderate, and the thin constituent garnet layers used within MPC would have been almost visually clear. This can be demonstrated by entering a design string such as SM[68.39] into design formula textbox, and running absorption-mode calculation, (e.g. using $\alpha$ = 2000 cm$^{-1}$), to reveal by zooming the graph area with mouse, that individual MO layers absorb only about 1.2% of the incident power on each single-pass transmission).

### 2.2. Generating optimised antireflector film designs using spectral target points

It is possible to apply additional optimisation constraints at up to three selected wavelength points, to force the algorithm to output the designs with specific spectral features, in either the transmission or reflection mode. An example of obtaining the optimised 5-layer thin-film broadband antireflector coating designs is stored in file Try_optimise_5-layer antireflector.mpc which is placed into the subdirectory of "Optimisation results files" in the installation directory. Menu item "Extra Data | Multi-wavelength spectral targets" is used to enter the additional optimisation constraints, regardless of whether the Faraday rotation features are being optimised or not. Figure 6 shows the required inputs within the two submenus related to the multi-wavelength spectral targets and the Incidence Geometry options, which will result in reproducing the 5-layer antireflector design shown also in Fig.

6. Selecting the design with maximum transmission at the design wavelength (either after running the optimisation, or simply after loading the relevant example file) will reveal the reflection spectrum as shown.

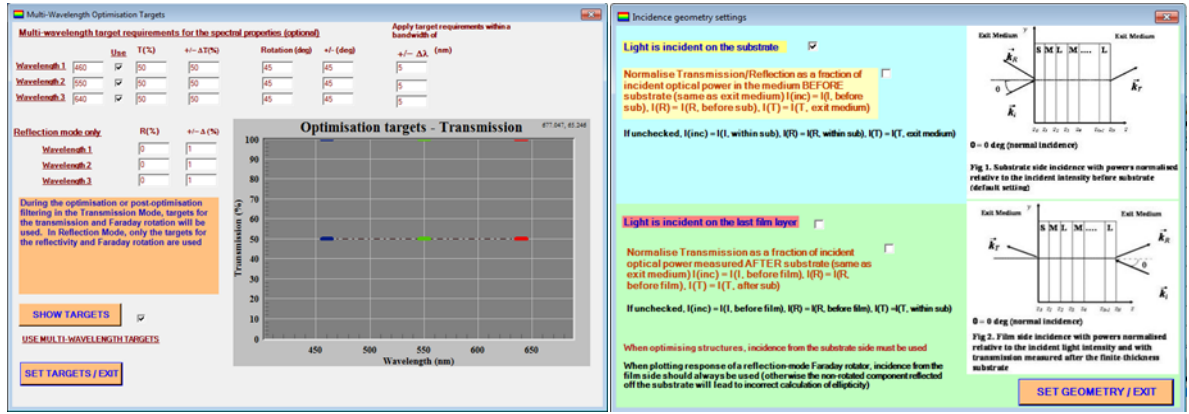

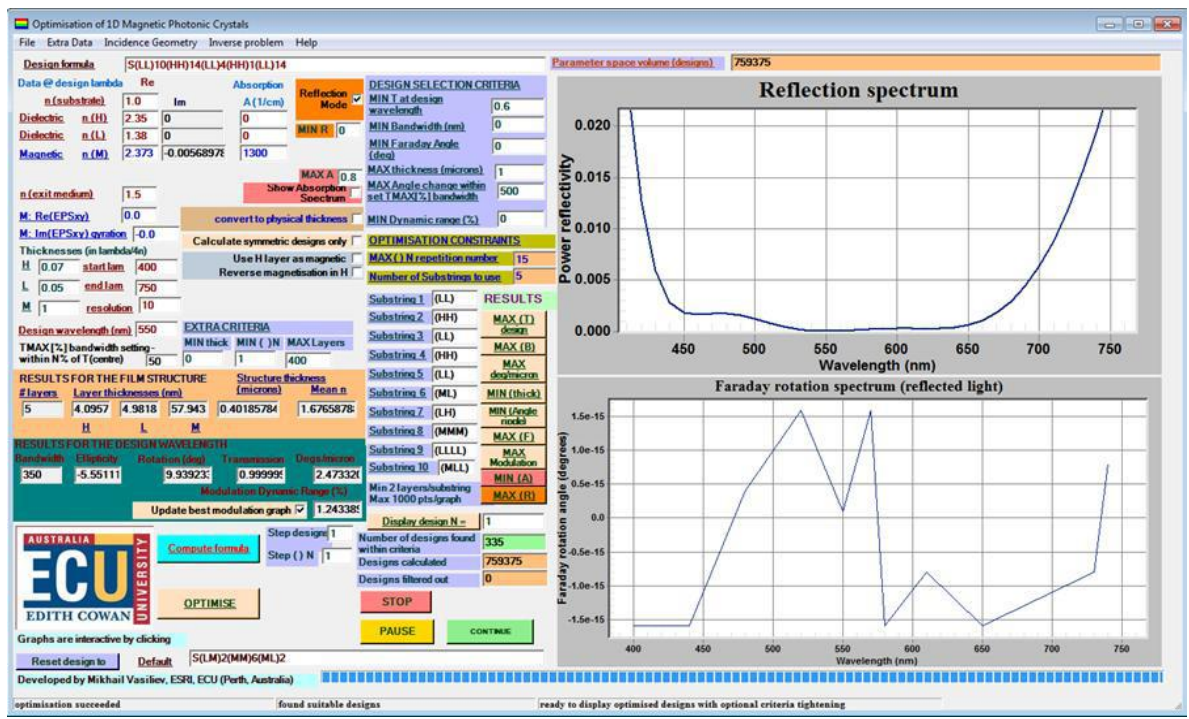

**Fig. 6.** Example of menu and algorithm settings required to generate a number of optimised 5-layer antireflector film designs on a glass substrate. The optical materials (ZnS and MgF$_2$) are presumed to possess constant refractive indices and zero absorption across the entire visible spectrum in this example.

This example also demonstrates the use of scaled QWOT layers for use in thin substrings, the thickness of which is then being optimised by the algorithm by adjusting the repetition indices. The calculation of more than 750000 designs should still take only a few minutes. A number of antireflector-type designs can be revealed by using the button "display design N=" with any corresponding number not exceeding the number of designs found within criteria.

Since the optimisation algorithm presumes the substrate-side incidence, it is convenient to define air as subtrate, and set the exit medium to glass. The obtained design S(LL)10(HH)14(LL)4(HH)1(LL)14 needs to have its deposition sequence reversed; and be re-evaluated for the film-side incidence case from air; using the physical thickness notation is preferable in this case, i.e. the design needs to be changed into S(LL)14(HH)1(LL)4(HH)14(LL)10.

In physical thickness notation, for a practical deposition-ready design description, this is equivalent to SL[139.49]H[8.19]L[39.85]H[114.68]L[99.64]. It is important to not forget to set n(S) to 1.5 and n(exit) to 1.0

(air) in this case. The incidence geometry settings can now be checked to correspond to the film-side incidence. Note the way the reflectivity of the back side of the substrate is accounted for in the detailed incidence geometry settings.

*2.3. Fitting of the measured thin-film transmission spectra to theoretical models*

One of the most frequently used applications of MPC Optimisation software, apart from optimising the MPC structures, is expected to be the fitting of actual deposited layer thickness, for thin-film materials with known refractive index dispersion function. The option of loading the measured transmission spectrum for immediate comparison with the modelled transmission spectrum of the same substrate-film system is available from menu item Inverse problem | Load T spectrum for fitting. To enable accurate modeling, both of the two checkboxes corresponding to the substrate-side incidence within menu Incidence Geometry | Define geometry must be checked, which is done by default. A measured reflection spectrum (if available for the normal  incidence reflection, which is rare with most instruments) can also be fitted in the same way as transmission, by running the calculations in reflection mode, with the corresponding checkbox checked. During the calibration of in-situ thickness control systems e.g. quartz crystal microbalances, or reflectometer-type systems, the task of determining the actual physical thickness of deposited thin-film layers is very common, and that is where MPC Optimisation software can be used effectively, in conjunction with other methods such as SEM or profilometry characterisation. Figure 6 illustrates graphically the results of fitting the loaded (measured) transmission spectra of two thin films of composition type $Bi_{0.9}Lu_{1.85}Y_{0.25}Fe_{4.0}Ga_1O_{12}$ (the refractive index dispersion data file for this composition is supplied within the appropriate sub-folder in the program installation directory of the target machines). A thinner (684 nm) film modeled as deposited on a glass substrate (n(S) = 1.5) was fitted using the refractive index and absorption coefficient data file related to the as-deposited (amorphous-phase) garnet films of this composition. Since the transmission of an annealed (garnet-phase) film was actually loaded, its transmission at shorter wavelengths was in excess of that modeled; the absorption of garnet-phase films is  practically always less in the crystallized state, compared to amorphous garnet-precursor films. A thicker film of fitted thickness (transmission shown in Fig. 7(b)) was measured in the amorphous phase – and therefore the quality of fit is better. Some systematic transmission discrepancies across a wide spectral range can be attributed to a combination of possible light scattering on film surface features and film layer refractive-index non-uniformity.

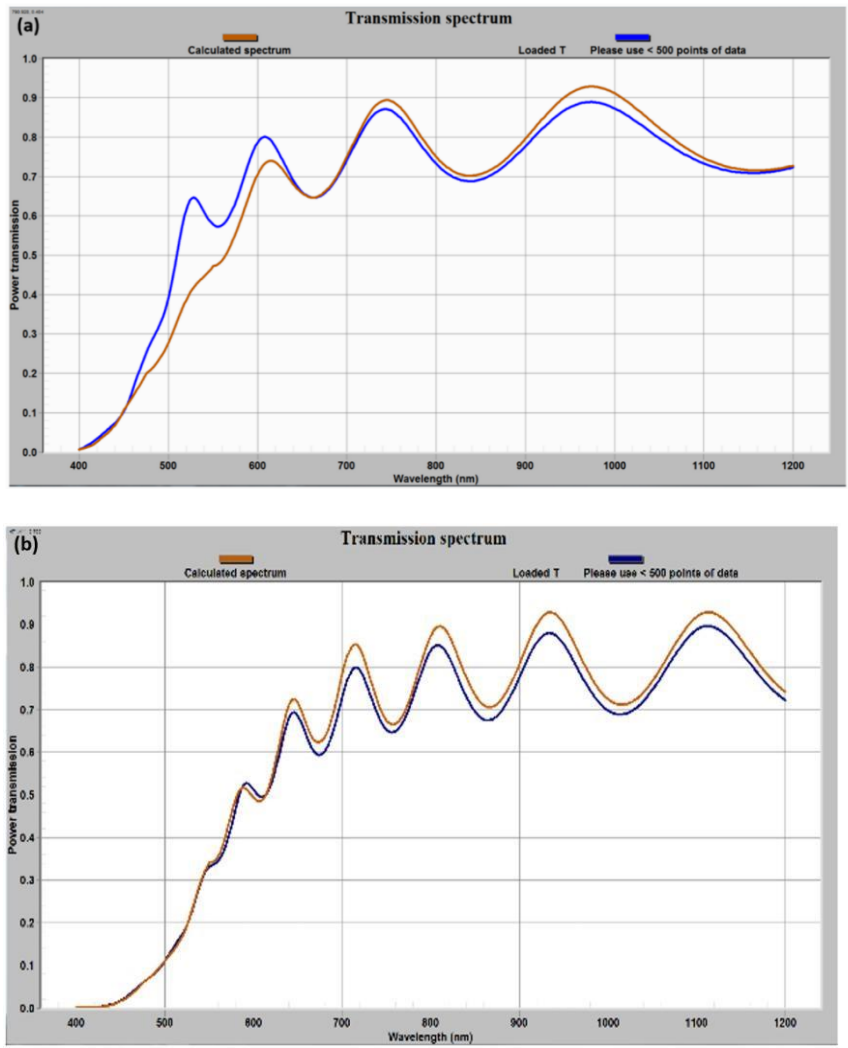

**Fig. 7.** Magneto-photonic crystal (MPC) software fitted transmission spectra of different thin film garnet layers; (a) annealed garnet sample of composition type $Bi_{0.9}Lu_{1.85}Y_{0.25}Fe_{4.0}Ga_{1}O_{12}$ and thickness 684 nm, (b) as-deposited garnet sample of the same composition but from another batch, of thickness 1310 nm.

If the spectral dependency of the film material absorption coefficient is completely unknown (or the data are not reliable), but the dispersion of its refractive index is well-known (e.g. from variable-angle spectroscopic ellipsometry data), then the index dispersion data files need to be prepared with zero values entered for all absorption coefficients. These data file still enable, in many cases, very reliable fitting of the physical thickness. It is highly recommended to then back up these thickness fitting results by also applying Swanepoel method-based techniques e.g. methods reported in [23].

*2.4. Fitting of the absorption coefficient spectra in single-layer films of known refractive index dispersion function, transmission spectrum, and thickness*

In situations when only the refractive index spectral distribution and the measured transmission spectrum of a semitransparent material layer are available, it is possible to use the custom-prepared "zero absorption" refractive index dispersion data file, and then first fit the physical thickness (typical results are shown in Fig. 8(a)), followed by a derivation of fitted absorption coefficient spectrum using a dedicated algorithm from program menu item Inverse problem | Derive absorption spectrum. In order to minimize errors, this combination of fitting procedures is only recommended, if several deposited films of the same material are available, having a significantly different physical thickness, and the fitting procedures applied to more than one film consistently lead to obtaining well-correlated datasets as a result of absorption fitting. It is also recommended to apply Swaneopol method-based fitting procedures [23], to re-confirm the validity and accuracy of the physical thickness data.

Figure 8 illustrates the results of physical thickness and absorption fitting procedures obtained using a transmission file of a 481 nm thick $Bi_{0.9}Lu_{1.85}Y_{0.25}Fe_{4.0}Ga_1O_{12}$ as-deposited film on a glass substrate, and the corresponding material's "zero-absorption" refractive index data file, both of which are included with program distribution into the corresponding subdirectories of the installation folder. The data file named 701 nm $Bi_{0.9}Lu_{1.85}Y_{0.25}Fe_4Ga_1O_{12}$ garnet layer_n (SWEM) at A=0.txt must be loaded using "load M data" button and the subfolder corresponding to this material type (we previously characterised the refractive index of a 701 nm thick film of this material). After the best fitted thickness value (481 nm) is obtained by comparing different models of design string, such as SM[450], ….etc. SM[481], the same design string must be re-entered using a quarterwave thickness notation, e.g. S(MM)3, where the QWOT multiplier for M thickness is set to 1.125 (for n(M) = 2.21, corresponding to the nearest index data point to the default 630 nm), to match the physical thickness of 481 nm in this notation. The textbox "structure thickness (microns)" must be used to check the physical thickness changes in response to changing either the (MM)N repetition index, or the QWOT multiplier for M layers. The index n(M) = 2.21 should be entered into its corresponding textbox, after looking up the 630 nm (default design wavelength) data for the actual refractive index, to avoid possible misrepresentations of QWOT data. This notation conversion is required for running the absorption coefficient fitting algorithm, as well as using M-type layers only – regardless of whether the material possesses any magnetic properties or not.

Once within the "derive absorption spectrum" submenu, the same measured transmission file should again be re-loaded onto the graph from file, followed by a relatively self-explanatory procedure for deriving the graph of absorption spectrum. No changes are usually required in any other textboxes. The result of fitting procedure with the above-described data files is shown in Fig 8(b).

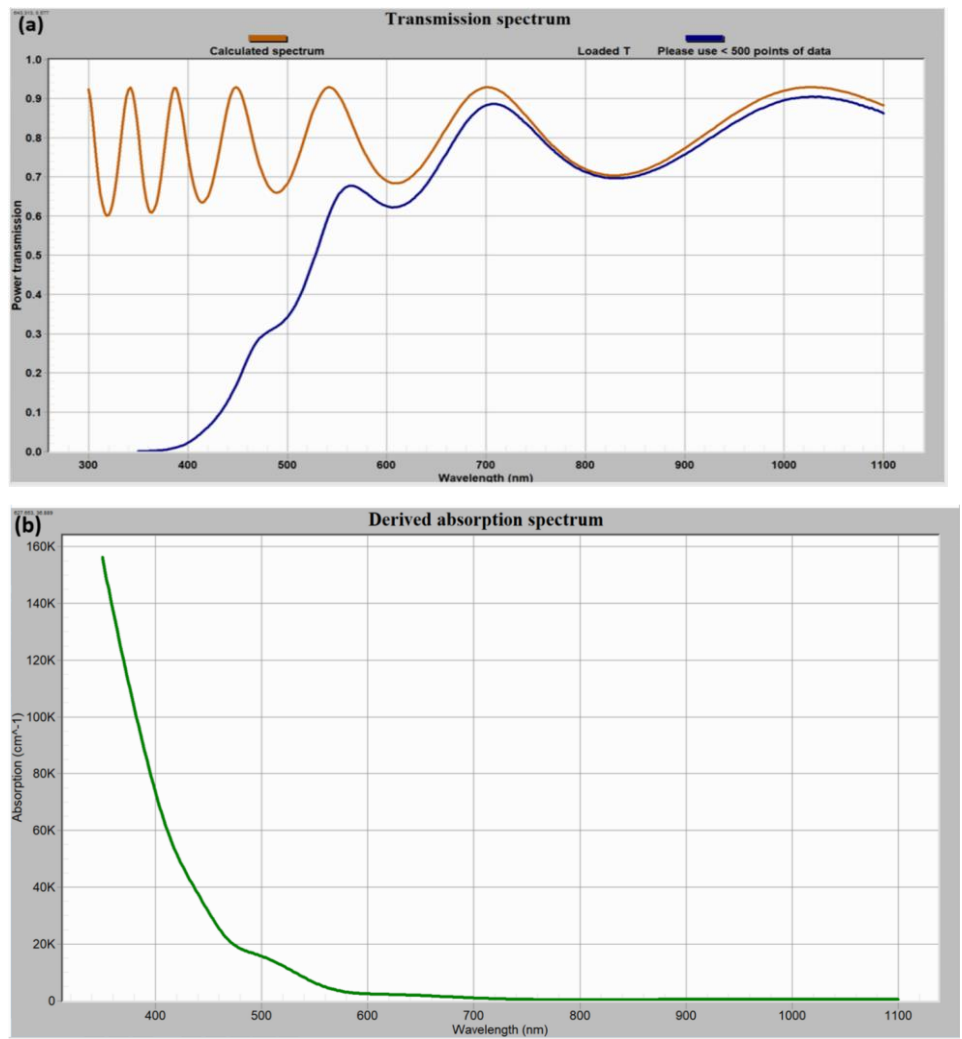

**Fig. 8.** Iterative (bisection algorithm-assisted) fitting of the absorption coefficient spectral dependency and the required pre-fitting of film thickness through matching transmission peak features. **(a)** Peak-aligned transmission spectrum pre-fitting result, from which the physical film thickness information and the measured dataset on the refractive index dispersion are then used, within the option available in the "Inverse Problem" menu, to derive the absorption coefficient data; **(b)** the algorithm-derived (fitted) absorption coefficient spectrum ($\alpha$, cm$^{-1}$) for 481nm-thick as-deposited $Bi_{0.9}Lu_{1.85}Y_{0.25}Fe_{4.0}Ga_1O_{12}$ garnet-precursor film sample on a glass substrate (its measured transmission spectrum is shown as blue curve in part (a)).

In the cases when the fitting algorithm produces exceptions such as described within a dialog box text, e.g. " At some or all wavelengths, even at zero absorption in M layers, the transmission of this structure should be less than specified", adjustments can be made to either check that the incidence geometry settings are correct, or the refractive index of the substrate can be increased in the model, removing these fitting procedure errors. In cases when these issues persist in the low-absorption wavelength ranges, the wavelength range of the model can be reduced to include only the regions where the fit results can be obtained. Non-uniformities in real thin films can lead to reduced refractive index values, leading then to reductions in the modeled reflection, thus showing increased transmission at some wavelengths compared to theory models.

After the absorption spectrum fitting procedure has been completed, the plotted data points can be exported into other formats, or saved in the data files using the options provided.

## 3. Installation

Installation of MPC Optimisation software is easy, and enabled by running the installer (.msi) file supplied within the .zip archived folder used for program redistribution. A necessary pre-requisite to program installation is Microsoft .NET Framework 1.1, which must be installed onto any Windows machine prior to running the MPC Optimisation installer. The .NET 1.1 Framework installation file (dotnetfx.exe) is supplied within the archived folder file used to redistribute MPC Optimisation. No known problems have so far been identified to exist in relation to installing this older version of .NET Framework onto modern computer systems. A number of example data files containing the samples of measured garnet thin film transmission spectra and files containing the data on the refractive index and absorption coefficient spectra of various garnets and other optical materials are placed into the selected program installation folder on installation. These data enable the users to re-create the example calculations presented within this article, and therefore are useful in mastering the software operation.

## 4. Code-related information

The program has been written as Microsoft Visual Studio .NET 2003 Solution, composed of three Projects: (i) Class Library project written using Managed C++ code, and built as .dll class library providing matrix-related computation functions; (ii) main program project, written using Visual C# and implemented as Windows Forms project; and (iii) deployment project, used for including the necessary reference assemblies, class libraries, license files, organising the file system structure within the installation directory, and generating the Installer files for program distribution. Installed MPC Optimisation program can run on any Windows system architecture, whether 32-bit or 64-bit, regardless of processor type.

Two main third-party software components have been embedded into the program structure, for which the necessary developer licenses have been purchased commercially. Bluebit Matrix Library 2.2 has been used to provide classes that enable efficient complex-valued 4x4 matrix operations functionality. The Class Library .dll assembly referencing the Bluebit Matrix Library assembly has been required, through the terms of developer grade matrix software license, to be compiled as signed assembly, using a strong-name assembly key file, which is not allowed for re-distribution by the developers. The second embedded component was GigaSoft ProEssentials 5 .NET package, which has been installed on the developer's computer to provide ActiveX controls that enable scientific-type graphing output functions. The assemblies GigaSoft.ProEssentials.dll and Pegrpcl.dll are referenced by the C# project, embedding the scientific graphing controls as Microsoft Component Object Model (COM) objects into the structure of Windows Forms project-related software assemblies. The use of licensed third-party software components, with the associated restrictions, is the primary reason why MPC Optimisation program code is not intended for open-source distribution. Additionally, newer versions of Bluebit Matrix Library designed to work with present-day versions of .NET Framework, as well as significant code syntax changes applied throughout the projects, would have been required to successfully port the Solution into current versions of Visual Studio, e.g. VS2015 or VS2017.

## 5. Conclusions

In summary, a software package has been described, designed to enable the design and optimization of 1D magnetic photonic crystals in terms of the achievable combinations of Faraday rotation, transmission and reflection spectra. The same package allows computational modelling of the optical spectral properties of various dielectrics-based generic single- and multilayer thin films. Additional program features include the tools for fitting of the experimentally-measured transmission or reflection spectra to theoretical models, allowing the film physical thickness data recovery, if detailed refractive index information is available. Fitting of the absorption coefficient spectra in absorbing material layers is possible, using an automated algorithm reliant on the data for the measured transmission spectrum, refractive index spectrum, and physical thickness.

**Notes**

The authors declare no competing financial interest.

**Acknowledgments**

This work was supported by Edith Cowan University.